\documentclass[a4paper,twoside]{article}
\usepackage{amssymb}
\usepackage{amsmath}
\usepackage{amsfonts}
\usepackage{graphicx}

\begin{document}

\title{\Large\bf{The Kantowski-Sachs Space-Time in Loop Quantum Gravity}}

\author{\\ Leonardo
 Modesto
 \\[1mm]
 \em\small{Centre de Physique Th\'eorique de Luminy,}\\ 
 \small{Universit\'e de la M\'editerran\'ee, F-13288 Marseille, EU}\\[-1mm]
  }

\date{\ } 
\maketitle

\begin{abstract}
We extend the ideas introduced in the previous work to a more general space-time. In particular 
we consider the Kantowski-Sachs space time with space section with topology $R \times S^2$. 
In this way we want to study a general space time that we think to be the space time inside the horizon of a black hole. In this case the phase space is four dimensional and we simply apply the quantization procedure suggested by Loop Quantum Gravity and based on an alternative to the Schroedinger representation introduced by H. Halvorson.  Through this quantization procedure we show that the inverse of the volume density and the Schwarzschild curvature invariant are upper bounded and so the space time is singularity free. Also in this case we can extend dynamically the space time beyond the classical singularity. 

\end{abstract}
 
\section*{Introduction}

This work is a generalization of the recent results obtained for the Schwarzschild solution inside the horizon and near the singularity where the operator $1/r$ and so the curvature invariant 
$\mathcal{R}_{\mu \nu \rho \sigma} \, \mathcal{R}^{\mu \nu \rho \sigma} = 48 M^2 G_N^2/r^6$
are non divergent in the quantum theory. This work is suggested from
a paper on Loop Quantum Cosmology \cite{Boj}. 
In this paper we use the same non Schr$\ddot{\mbox{o}}$dinger procedure of quantization used 
in the previous paper \cite{work1} and in the work of  V. Husain and O. Winkler on quantum cosmology but introduced by Halvorson \cite{Fonte.Math} and also by A. Ashtekar, S. Fairhust and
J. Willis \cite{AFW}. 

In this paper we focus on a general two dimensional minisuperspace with space section of topology $\mathbf{R} \times \mathbf{S}^2$ which is know as Kantowski-Sachs space time 
\cite{KS}. The Schwarzschild space time inside the horizon is a particular representative of this class of metrics. Using this method \cite{Thie} we can define 
the inverse volume density and the Schwarzschild curvature invariant in term of the holonomy analog and the volume itself and we show that these quantity are finite and upper bounded. Using also the result in \cite{Thie} we can obtain the Hamiltonian constraint in terms of
 the volume and at the quantum level we have a discrete equation depending on two 
 parameters for the coefficients of the physical states.

The paper is organized as follows : in the first section we report the metric we want to study.
we consider this space time as the interior of a black hole in a more 
general form respect to the work \cite{work1}. We calculate the hamiltonian constraint, the volume 
operator and we introduce the fundamental variables of the theory.

 In the third section we quantize the system using the non Schr$\ddot{\mbox{o}}$dinger procedure of quantization \cite{Fonte.Math}, \cite{Fonte} and \cite{AFW}.
In this section we show that the inverse volume operator and the curvature invariant are
singularity free in quantum gravity and also that the Hamiltonian constraint acts like a 
difference operator as in loop quantum cosmology.

\section{The Spherically Symmetric Space-Time}

We want to study a generical metric for an homogeneous, anisotropic space with spatial section of topology $\mathbf{R} \times \mathbf{S}^2$, this is the Kantowski-Sachs Space-Time. In this case we have two independent functions of
the time $a(t)$ and $b(t)$ and the metric assume the following form 
\begin{eqnarray}
ds^2 = - dt^2 + a^2 (t) dr^2 + b^2 (t)  (\sin^2 \theta d\phi^2 + d \theta^2).
\label{metricab}
\end{eqnarray}
In the previous paper \cite{work1} we considered the Schwarzschild solution, so $a(t)$  was not a general function of $t$ but it was a function of $b(t)$ which was the only independent function. 

The $Diff$-constraint for the class of metrics in (\ref{metricab}) is identically satisfied and the Hamiltonian constraint in terms of $\dot{a}$ and $\dot{b}$ is
 \begin{eqnarray}
H_L = |a| \, \dot{b}^2 + 2 \, \dot{a}  \, \dot{b} \, b \, \mbox{sgn}(a) + |a| ,
\label{HL.Mini} 
\end{eqnarray}
in terms of $p_a$ and $p_b$ is 
 \begin{eqnarray}
H_c =\frac{ G_N \, |a| \, p_a^2}{2R \, b^2} - \frac{G_N p_a p_b \, \mbox{sgn}(a)}{ R b} - \frac{R}{2 G_N} |a| .
\label{HC.Mini} 
\end{eqnarray}
The volume of a space section is
\begin{eqnarray}
V = \int dr \, d \phi \, d \theta \, h^{1/2} = 4 \pi R |a| b^2 .
\label{Volume}
\end{eqnarray}
where $R$ is a cutt - off on the space radial coordinate. We can work also with radial densities 
because the model is homogeneous and all the following results remain identical. In another way, 
the spatial homogeneity enable us to fix a linear radial cell $\mathcal{L}_r$ and restrict all 
integrations to this cell \cite{Boj}.   
We have two canonical pairs, one is given by $a \equiv x_a$ and $p_a$, the other is given by
$b \equiv x_b$ and $p_b$ for which the Poisson brackets are $\{x_a,p_a\} = 1$ 
and $\{x_b,p_b\} = 1$.
From now on we consider $x_a, x_b \in \mathbb{R}$ and we will introduce the modulus of $x_a$ and $x_b$ where it is necessary. This choice to take $x_a, x_b \in \mathbb{R}$ is not correct classically because we have a singularity in $b=0$, but the situation can be (a priori) different in quantum theory; and it will be, as we will see. 

Following \cite{work1} we introduce an algebra of classical observable and we write the quantities of physical interest in terms of these variable. As in Loop Quantum Gravity we use the fundamental variables $x_a, x_b$ and 
\begin{eqnarray}
&& U_{\gamma_a}(p) \equiv \mbox{exp} \Big(\frac{8 \pi G_N \gamma_a}{L_a^2} \, i \,  p_a\Big),  \nonumber \\
&& U_{\gamma_b}(p) \equiv \mbox{exp} \Big(\frac{8 \pi G_N \gamma_b}{L_b} \, i \,  p_b\Big), 
\label{UaUb}
\end{eqnarray}   
where $\gamma$ is a real parameter and $L$ fixes the unit of length. The parameter $\gamma$ is necessary to separate the momentum point in the phase space. This variable can be seen as the momentum analog of the holonomy variable of loop quantum gravity.\\
We have also that 
\begin{eqnarray}
&& \{x_a , U_{\gamma_a}(p_a)\} = 8 \pi G_N \frac{i \, \gamma_a}{L_a^2} U_{\gamma_a}(p_a), \nonumber \\
&& \{x_b , U_{\gamma_b}(p_b)\} = 8 \pi G_N \frac{i \, \gamma_b}{L_b} U_{\gamma_b}(p_b), \nonumber \\
&& U_{\gamma_a}^{-1} \{ V^m , U_{\gamma_a} \} = (4 \pi R |x_b|^2)^m \, m \, |x_a|^{m-1} \mbox{i} \gamma_a \frac{8 \pi G_N}{L_a^2} \mbox{sgn}(x_a), \nonumber \\
&& U_{\gamma_b}^{-1} \{ V^n , U_{\gamma_b} \} = (4 \pi R |x_a|)^n \, 2 n \, |x_b|^{2n-1} \mbox{i} \gamma_b \frac{8 \pi G_N}{L_b} \mbox{sgn}(x_b).
\label{Poisson.Volume}
\end{eqnarray}
From those relations we can construct the following quantities that we use extensively 
\begin{eqnarray}
&& \frac{|x_b|^{2/3}}{|x_a|^{2/3}} = - \frac{3 \, \mbox{i} \, L_a^2}{(4 \pi R)^{\frac{1}{3}} \, 8 \pi G_N \gamma_a} \, U_{\gamma_a}^{-1} \{ V^{\frac{1}{3} }, U_{\gamma_a} \} \, \mbox{sgn}(x_a), \nonumber \\
&& \frac{|x_a|^{1/4}}{|x_b|^{1/2}} = - \frac{2 \, \mbox{i} \, L_b}{(4 \pi R)^{\frac{1}{4}} \, 8 \pi G_N \gamma_b} \, U_{\gamma_b}^{-1} \{ V^{\frac{1}{4} }, U_{\gamma_b} \} \, \mbox{sgn}(x_b), \nonumber \\
&& \sqrt{|x_a|} = - \frac{ \mbox{i} \, L_b}{(4 \pi R)^{\frac{1}{2}} \, 8 \pi G_N \gamma_b} \, U_{\gamma_b}^{-1} \{ V^{\frac{1}{2} }, U_{\gamma_b} \} \, \mbox{sgn}(x_b), \nonumber \\
&& \frac{|x_a|^{1/3}}{|x_b|^{1/3}} = - \frac{3 \, \mbox{i} \, L_b}{2 \, (4 \pi R)^{\frac{1}{3}} \, 8 \pi G_N \gamma_b} \, U_{\gamma_b}^{-1} \{ V^{\frac{1}{3} }, U_{\gamma_b} \} \, \mbox{sgn}(x_b). \nonumber \\
\label{bsua}
\end{eqnarray}
We use this relation in the next section into the physical quantities. 
We are interested to the quantity
$\frac{1}{V}$ because classically this quantity can diverge as in the case of the Schwarzschild 
solution and can produce a singularity. The other very important operator we will consider is 
$1/|b|^6$ that corresponds 
to the curvature invariant $R_{\mu \nu \rho \sigma} R^{\mu \nu \rho \sigma}$ for the
Schwarzschild solution.  We are also interested to the Hamiltonian constraint and to the dynamics of the minisuperspace model.

\section{Quantum Theory}

We construct the quantum theory in analogy with the procedure used in loop quantum gravity and in particular following \cite{work1} but with two copy of canonical variable. First of all for any 
canonical couple $(x_a, p_a)$ or $(x_b, p_b)$ we must take an algebra of classical functions that is represented as quantum configuration operators. We choose the algebra generated by the function
\begin{eqnarray}
W(\lambda) = e^{i \lambda x/L}
\end{eqnarray}
where $\lambda \in \mathbb{R}$. The algebra consists of all function of the form 
\begin{eqnarray}
f(x) = \sum_{j=1}^{n} c_j e^{i \lambda_j x/L}
\end{eqnarray}
where $c_j \in \mathbb{C} $ and their limits with respect to the supremum norm and with $x$ we
indicate $x_a$ or $x_b$. This algebra is the $algebra$ $of$ $almost$ $periodic$ $function$ $over$ $\mathbb{R}$ ($AP(\mathbb{R})$). The algebra ($AP(\mathbb{R})$) is isomorphic to $C(\bar{\mathbb{R}}_{Bohr})$ that is the algebra of continuous functions on the Bohr-compactification of $\mathbb{R}$. This suggests to take the Hilbert space $L_2(\bar{\mathbb{R}}_{Bohr}, d \mu_0)$, where $d \mu_0$ is the Haar measure on $\bar{\mathbb{R}}_{Bohr}$.\\
With this choice the basis states in the complete Hilbert space are the tensor product  
\begin{eqnarray}
&& |\lambda_a \rangle \otimes  |\lambda_b \rangle \equiv |e^{i \lambda_a x_a} \rangle \otimes
       |e^{i \lambda_b x_b/L_b} \rangle, \nonumber \\
&& \langle \mu_a | \lambda_a \rangle = \delta_{\mu_a, \lambda_a} \hspace{1cm} 
      \langle \mu_b | \lambda_b \rangle = \delta_{\mu_b, \lambda_b}. 
      \end{eqnarray}
The action of the configuration operators $\hat{W_a}(\lambda_a)$ and $\hat{W_b}(\lambda_b)$ on the base is definited by 
\begin{eqnarray}
&& \widehat{W}_a(\lambda_a) | \mu_a \rangle = e^{i \lambda_a \hat{x}_a} | \mu_a \rangle = e^{i \lambda_a \mu_a } |\mu_a \rangle, \nonumber \\
&& \widehat{W}_b(\lambda_b) | \mu_b \rangle = e^{i \lambda_b \hat{x}_b/L_b} | \mu_b \rangle = e^{i \lambda_b \mu_b } |\mu_b \rangle.
\end{eqnarray}
Those operators are weakly continuous in $\lambda_a$,  $\lambda_b$ and this imply the existence of  self-adjoint operators $\hat{x}_a$ and $\hat{x}_b$, acting on the basis states according to  
\begin{eqnarray}
&& \hat{x}_a |\mu_a \rangle = \mu_a |\mu_a \rangle,  \nonumber \\
&& \hat{x}_b |\mu_b \rangle = L_b \mu_b |\mu_b \rangle.
\label{xoperator}
\end{eqnarray}
Now we introduce the operators corresponding to the classical momentum functions 
$U_{\gamma_a}$ and $U_{\gamma_b}$ of (\ref{UaUb}).
 We define the action of $\hat{U}_{\gamma_a}$ and $\hat{U}_{\gamma_b}$ on the basis states using the definitions (\ref{xoperator}) and using a quantum analog of the Poisson brackets between $x_a$ and $U_{\gamma_a}$ and $x_b$ and $U_{\gamma_b}$ 
\begin{eqnarray}
&& \hat{U}_{\gamma_a} |\mu_a \rangle = | \mu_a - \gamma_a \rangle \hspace{1cm} 
      \hat{U}_{\gamma_b} |\mu_b \rangle = | \mu_b - \gamma_b \rangle,    
 \nonumber \\ 
&& \left[ \hat{x}_a , \hat{U}_{\gamma_a} \right] = - \gamma_a \hat{U}_{\gamma_a} \hspace{1.5cm}
       \left[ \hat{x}_b , \hat{U}_{\gamma_b} \right] = - \gamma_b \, L_b \, \hat{U}_{\gamma_b}.
       \end{eqnarray} 
Using the standard quantization procedure $[ \, , \, ] \rightarrow i \hbar \{ \, , \, \}$, and using the the
first two equations of (\ref{Poisson.Volume}) we obtain 
\begin{eqnarray}
&& L \equiv L_a = L_b = \sqrt{8 \pi G_n \hbar}.
\end{eqnarray}

\subsection{Non singular space-time
}

We use equation (\ref{Volume}) and in particular the form $V = 4 \pi R |x_a| \, |x_b|^2$. So the action of the volume operator on the basis states is 
\begin{eqnarray}
\hat{V} | \mu, \nu \rangle = 4 \pi R \, |\hat{x}_a| \, |\hat{x}_b|^2 | \mu, \nu \rangle = 4 \pi R L_b^2  \, |\mu| \, |\nu|^2  |\mu, \nu \rangle.
\end{eqnarray}
Now we show that the operator $\frac{1}{\mbox{det(E)}} = \frac{1}{\sqrt{h}} \sim
 \frac{1}{|a| \, |b|^2}$ doesn't diverge in the quantum theory in the singularity point 
 $b =0$.\\
We use the relations (\ref{bsua}) and we promote the Poisson Brackets to commutators. In this way we obtain (for $\gamma_a = \gamma_b =1$) the operator 
\begin{eqnarray}
\widehat{\frac{1}{\mbox{det(E)}}} = \widehat{\Bigg(\frac{|x_a|}{\,\,\,|x_b|^2}\Bigg)^3}_{\gamma_b} \, 
                                                              \widehat{\Bigg(\frac{|x_b|^2}{|x_a|^2}\Bigg)^3}_{\gamma_a} \, 
                                                              \widehat{\Bigg(\frac{|x_a|}{|x_b|}\Bigg)^2}_{\gamma_b}, 
                                                              \label{1sudetE}
                                                           \end{eqnarray}
the action of this operator on the bases states is   
  \begin{eqnarray}
\widehat{\frac{1}{\mbox{det(E)}}} |\mu, \nu \rangle \, = \, \frac{2^6 \, 3^{15}}{L^2} \, 
                               |\mu|^5 \, |\nu|^6 \, [|\nu -1|^{\frac{1}{2}} -|\nu|^{\frac{1}{2}}]^{12} \, 
                               ||\mu -1|^{\frac{1}{3}} - |\mu|^{\frac{1}{3}}|^9 \, 
                               [|\nu -1|^{\frac{2}{3}} - |\nu|^{\frac{2}{3}}]^6.
 \end{eqnarray}
 
As we can see, the spectrum is bounded from below and so we have no singularity in 
the quantum theory in $b = 0$.

The other operator we want to study is $1/|b|^6$. This operator corresponds to
the curvature invariant $R_{\mu \nu \rho \sigma} R^{\mu \nu \rho \sigma} \sim \frac{1}{b^6}$
that diverges in $b=0$ in the classical Schwarzschild solution. 
To obtain information about the singularity
at the quantum level, we consider the operator $\widehat{1/|b|}$. 

Using the relations 
(\ref{bsua}), we can define the operator $\widehat{1/|b|}$ as 
\begin{eqnarray}
\widehat{\frac{1}{|b|}} = \widehat{\Bigg(\frac{|x_a|}{\,\,\,|x_b|^2}\Bigg)}_{\gamma_b} \, 
                                                              \widehat{\Bigg(\frac{|x_b|^2}{|x_a|^2}\Bigg)}_{\gamma_a} \, 
                                                              \widehat{\Bigg(\frac{|x_a|}{|x_b|}\Bigg)}_{\gamma_b}.
                                                              \label{1sub}
\end{eqnarray}
The operator $\widehat{1/|b|}$ 
is diagonal on the basis states and the spectrum (for $\gamma_a = \gamma_b =1$) is 
\begin{eqnarray}
\widehat{\frac{1}{|b|}} |\mu, \nu \rangle = \frac{2 \, 3^6}{L} \, 
|\mu|^2 \, \Big| |\mu -1|^{\frac{1}{3}} - |\mu|^{\frac{1}{3}} \Big|^3 \, 
|\nu|^2  \, 
\left[ |\nu -1|^{\frac{1}{2}} - |\nu|^{\frac{1}{2}} \right]^4 \, 
\Big|\nu -1|^{\frac{2}{3}} - |\nu|^{\frac{2}{3}} \Big|^3 \, 
|\mu, \nu \rangle 
\label{1subspec}
\end{eqnarray}
This operator doesn't diverge in $\nu = 0$ (or $b = 0$), where the classical singularity is localized. In particular from the Schwartzschild solution we know that the 
singular point is in $b = 0$ and $a = \infty$, so if we take the limit $ \mu \rightarrow \infty$ in the
spectrum we obtain 
\begin{eqnarray}
\widehat{\frac{1}{|b|}} |\mu, \nu \rangle \rightarrow \frac{2 \, 3^3}{L} \, 
|\nu|^2  \, 
\left[ |\nu -1|^{\frac{1}{2}} - |\nu|^{\frac{1}{2}} \right]^4 \, 
\Big|\nu -1|^{\frac{2}{3}} - |\nu|^{\frac{2}{3}} \Big|^3 \, 
|\mu, \nu \rangle.
\label{1subspeclim}
\end{eqnarray}
The spectrum now depends only on the eigenvalues $\nu$ and 
it is singularity free in $\nu = 0$ (and $\mu \rightarrow \infty$).

We can conclude that the Schwarzschild curvature invariant 
$R_{\mu \nu \rho \sigma} R^{\mu \nu \rho \sigma} \sim \frac{1}{b^6}$ is
not divergent in quantum gravity and we can extend the space-time 
beyond the classical singularity in $r \equiv b =0$.






\subsection{Hamiltonian Constraint}

We said that the Hamiltonian for our system depend on two canonical couples and we 
report now this constraint 
\begin{eqnarray}
H_c  = \frac{G_N \, p_a^2}{2 R} \, \frac{|x_a|}{\,\, x_b^2}
            - \frac{G_N \, p_a \, p_b}{R} \, \frac{\mbox{sgn}(x_b) \, \mbox{sgn}(x_a)}{|x_b|} 
            - \frac{R}{2 G_N} \, |x_a|
\label{Hamiltonian.1}
\end{eqnarray}
Now we quantize this Hamiltonian constraint. As we know, the operators $p_a$ 
and $p_b$ don't exist in our
quantum representation and so we choose the following alternative representation for the operators $p_a^2$ and $p_a \, p_b$.
We start from the classical expressions
\begin{eqnarray}
&& p_a^2 = \frac{L_a^4}{(8 \pi G_N)^2} \mbox{lim}_{\gamma_a \rightarrow 0} \left( \frac{2 - U_{\gamma_a} - U_{\gamma_a}^{-1}}{\gamma_a^2} \right), \nonumber \\
&& p_a \, p_b = \frac{L_a^2 \, L_b}{2(8 \pi G_N)^2} \, \mbox{lim}_{\gamma_a, \gamma_b \rightarrow 0} 
            \Bigg[ \Bigg(\frac{U_{\gamma_a} + U_{\gamma_b} - U_{\gamma_a} \, U_{\gamma_b}  -1}{\gamma_a \, \gamma_b}\Bigg) + \Bigg(\frac{U_{\gamma_a}^{-1} + U_{\gamma_b}^{-1} - U_{\gamma_a}^{-1} \, U_{\gamma_b}^{-1}  -1}{\gamma_a \, \gamma_b}\Bigg)\Bigg]. \nonumber \\
            &&
\end{eqnarray}
We have a physical interpretation setting $\gamma_a = \gamma_b = l_F / L_{phys}$, where $L_{Phys}$ is the characteristic size of the system and $l_F$ is a fundamental length scale. In our case $l_F = l_P$ and $\gamma_a = \gamma_b = l_P / L_{Phys}$. We are ready to write the Hamiltonian constraint 
\begin{eqnarray}
\hat{H} & = & \frac{1}{32 \pi^2 G_N R^2 \gamma_a^2 \gamma_b^4} \left[ 2 - \hat{U}_a -  \hat{U}_{a}^{-1} \right]  \, 
    \left( \hat{U_b}^{-1} \left[ \hat{V}^{\frac{1}{4}} , \hat{U_b} \right] \right)^4 \nonumber \\
    & + & \frac{3^6}{2^{11} \pi^5 R^4 L^4 G_N \gamma_a^7 \gamma_b^5} \,   \Bigg[\Big( \frac{\hat{U}_a + \hat{U}_b - \hat{U}_a \, \hat{U}_b  -1}{2}\Big)+ \Big( \frac{\hat{U}_a^{-1} + \hat{U}_b^{-1} - \hat{U}_a^{-1} \, \hat{U}_b^{-1}  -1}{2}\Big)\Bigg]  \nonumber \\
    && \hspace{3.4cm} \left( \hat{U_b}^{-1} \left[ \hat{V}^{\frac{1}{4}} , \hat{U_b} \right] \right)^4
    \, \left( \hat{U_a}^{-1} \left[ \hat{V}^{\frac{1}{3}} , \hat{U_a} \right] \right)^3 
    \, \left( \hat{U_b}^{-1} \left[ \hat{V}^{\frac{1}{3}} , \hat{U_b} \right] \right) ^3   + \nonumber \\
    & - & \frac{1}{8 \pi G_N L^2 \gamma_b^2} \left( \hat{U_b}^{-1} \left[ \hat{V}^{\frac{1}{2}} , \hat{U_b} \right]
     \right) ^2.
     \label{Hconstraint}
            \end{eqnarray}
Now we resolve the Hamiltonian constraint. The solutions of the Hamiltonian constraint are
 in the $\mathcal{C}^{\star}$ space that is the dual of the dense subspace $\mathcal{C}$ of the kinematical space $\mathcal{H}$. A generic 
element of this space is 
\begin{eqnarray}
\langle \psi | = \sum_{\mu, \nu} \psi(\mu, \nu) \langle \mu, \nu|.
\end{eqnarray}
The constraint equation $\hat{H} |\psi \rangle = 0$ is now interpreted as an equation in the dual space $\langle \psi | \hat{H}^{\dag}$;
from this equation we can derive a relation for the coefficients $\psi(\mu, \nu)$ 
\begin{eqnarray}
&& [2 \alpha(\mu, \nu) - 2 \beta(\mu, \nu) + \gamma(\mu, \nu)] \, \psi(\mu, \nu) -
[\alpha(\mu + \gamma_a, \nu) - \beta(\mu+\gamma_a, \nu)] \, \psi(\mu + \gamma_a, \nu) + \nonumber \\
&& \hspace{0cm} - [\alpha(\mu - \gamma_a, \nu) + \beta(\mu - \gamma_a, \nu)] \, \psi(\mu - \gamma_a, \nu) +
       \beta(\mu, \nu + \gamma_b) \,\psi(\mu, \nu + \gamma_b) + \nonumber \\
       && - \beta(\mu, \nu - \gamma_b) \, 
       \psi(\mu, \nu -\gamma_b) +
       \beta(\mu+\gamma_a, \nu + \gamma_b) \, \psi(\mu+\gamma_a, \nu+\gamma_b) \nonumber \\
       && - \beta(\mu - \gamma_a, \nu -\gamma_b) \, \psi(\mu - \gamma_a, \nu -\gamma_b)  = 0 
      \label{difference}
\end{eqnarray}
where the function $\alpha, \beta, \gamma$ are 
\begin{eqnarray}
 \alpha(\mu, \nu) & = &\frac{L^2}{8 \pi^2 R G_N \gamma_b^4 \gamma_a^2} \, 
      \Big(|\mu|^{\frac{1}{4}} |\nu - \gamma_b|^{\frac{1}{2}} -|\mu|^{\frac{1}{4}} |\nu|^{\frac{1}{2}} \Big)^2,  \nonumber \\
 \beta(\mu, \nu) & = & - \frac{L^2}{2(8 \pi)^2 G_N R \gamma_b^5 \gamma_a^7} \, 
       \Big(|\mu|^{\frac{1}{4}} |\nu - \gamma_b|^{\frac{1}{2}} -|\mu|^{\frac{1}{4}} |\nu|^{\frac{1}{2}} \Big)^4
       \Big(|\mu - \gamma_a|^{\frac{1}{3}} |\nu|^{\frac{2}{3}} -|\mu|^{\frac{1}{3}} |\nu|^{\frac{2}{3}} \Big)^3  \nonumber\\
     &&\hspace{2.8cm}\Big(|\mu|^{\frac{1}{3}} |\nu - \gamma_b|^{\frac{2}{3}} -|\mu|^{\frac{1}{3}} |\nu|^{\frac{2}{3}} \Big)^3 \nonumber \\        
     \gamma(\mu, \nu) & = & \frac{R}{2 G_N \gamma_b^2} \, 
     \Big(|\mu|^{\frac{1}{2}} |\nu - \gamma_b| - |\mu|^{\frac{1}{2}} |\nu| \Big)^2.     
\end{eqnarray}

The relation (\ref{difference}) determines the coefficients for the physical dual state and we can interpret this states as $quantum$ $space$ $time$ near the classical point $b=0$, which corresponds to the singularity of the space time in the case of classical black hole solution. 
From the difference equation (\ref{difference}) 
we obtain physical states as combinations of a countable number of  components of the form
$\psi(\mu + n \gamma_a, \nu + m \gamma_a) |\mu + n \gamma_a, \nu + \gamma_b \rangle $ ($\gamma_a \gamma_b \sim l_P/L_{Phys} \sim 1$); any component corresponds to a particular value of volume, so we can interpret $\psi(\mu + \gamma_a, \nu + \gamma_b)$
as the function of the Black Hole inside the horizon 
at the time $\nu + \gamma_b$ if we interprete $b$ as the time and $a$ as the space partial observable that defines the quantum fluctuations around the Schwarzschild solution.  
A solution of the Hamiltonian constraint corresponds to a linear combination of black hole 
states for particular  values of the times.

\section*{Conclusions}

In this work we have applied the quantization procedure of \cite{Fonte} to the Kantowski-Sachs space time \cite{KS} with space topology  $\mathbf{R} \times \mathbf{S}^2$. This space time contains the part of Schwarzschild solution on the other side of the horizon as a particular classical solution.
The quantization procedure is alternative to the Schr$\ddot{\mbox{o}}$dinger quantization and it is   
suggested by loop quantum cosmology. 

The main results are :

\begin{enumerate}
\item  the inverse volume operator has a finite spectrum near the point $b=0$; 
in particular the Schwarzschild curvature invariant  
$R_{\mu \nu \rho \sigma} R^{\mu \nu \rho \sigma} \sim \frac{1}{b^6}$ doesn't 
diverge for $b = 0$ in the quantum theory and we can conclude that the classical 
Schwarzschild black hole 
singularity disappears in quantum gravity,
\item  the solution of the Hamiltonian constraint gives a discrete difference equation for the coefficients of the physical states and we can have many scenarios to connect our universe  
to another. 
\end{enumerate}

\section*{Acknowledgements}
We are grateful to Carlo Rovelli and Eugenio Bianchi for many important and clarifying discussion about this work. This work is supported in part by a grant from the Fondazione Angelo Della Riccia.


\begin{thebibliography}{9}

\bibitem{book}
Carlo Rovelli, {\em Quantum Gravity}, (Cambridge University Press,
Cambridge, 2004)

\bibitem{work1} Leonardo Modesto, {\em Disappearance of the Black Hole Singularity in Quantum Gravity} ; \\gr-qc/0407097

\bibitem{Boj} M. Bojowald {\em Inverse scale factor in isotropic quantum geometry}, Phys. Rev. {\bf D64} 084018
(2001); M. Bojowald {\em Loop Quantum Cosmology \textrm{IV}: discrete time evolution}, 
Class. Quant. Grav. {\bf 18}, 1071 (2001); 
A. Ashtekar, M. Bojowald and J. Lewandowski, {\em Mathematica structure of loop quantum cosmology}, Class. Quant. Grav. (2003)
  
\bibitem{KS} R. Kantowski and R. K. Sachs, J. Math. Phys. {\bf 7} (3) (1966)

\bibitem{AFW} A. Ashtekar, S. Fairhurst and J. Willis, {\em Quantum gravity, shadow states, and quantum mechanics}, Class. Quant. Grav. {\bf 20} 1031-1062 (2003)

\bibitem{Partial} C Rovelli, ``Partial observables", {Phy
Rev} {D65} (2002) 124013; gr-qc/0110035

\bibitem{Fonte} Viqar Husain and Oliver Winkler, {\em On singularity resolution in quantum gravity} ; gr-qc/0312094

\bibitem{Fonte.Math} H. Halvorson, {\em Complementary of representations in quantum mechanics} ; quant - ph/0110102

\bibitem{Thie} T. Thiemann, {\em Quantum Spin Dynamics}, Class. Quant. Grav. {\bf 15}, 839 (1998) 

\bibitem{Thie.2} T. Thiemann, {\em Introduction to Modern Canonical Quantum General Relativity} ; gr -qc/0110034; {\em Lectures on Loop Quantum Gravity}, gr-qc/0210094

\end{thebibliography}
\end{document}